\documentclass{aa}

\usepackage[varg]{txfonts}
\usepackage{graphicx}
\usepackage{times}
\usepackage[figuresright]{rotating}
\usepackage{longtable,lscape}
\usepackage{amsmath}
\usepackage{natbib}
\bibpunct{(}{)}{;}{a}{}{,} 

\begin{document}

        \title{The supersoft X-ray source in \object{V5116 Sgr}}
        \subtitle{I. The high resolution spectra\thanks{Based on observations obtained with XMM-Newton, an ESA science mission with instruments and contributions directly funded by ESA Member States and NASA.}}

        \author{G.~Sala\inst{1,2}
          \and J.U.~Ness\inst{3}
          \and M.~Hernanz\inst{4,2}
          \and J.~Greiner\inst{5}
        }

        \institute{Departament de F\'isica, EEBE, Universitat Polit\`ecnica de Catalunya. BarcelonaTech. 
                Av. d'Eduard Maristany 10-14, 08019, Barcelona, Spain\\
        \email{gloria.sala@upc.edu}
        \and
        Institut d'Estudis Espacials de Catalunya, c/Gran Capit\`a 2-4, Ed. Nexus-201, 08034, Barcelona, Spain
        \and
        Science Operations Division, Science Operations Department of ESA, ESAC, 28691 Villanueva de la Cañada, Spain
        \and
        Institut de Ci\`encies de l'Espai (ICE-CSIC). Campus UAB. c/ Can Magrans s/n, 08193 Bellaterra, Spain
        \and
        Max-Planck-Institut f\"ur Extraterrestrische Physik, Giessenbach-Str. 1, D-85748 Garching, Germany        
        }

        \offprints{G. Sala}

        \date{Received ... /accepted ...}

 
\abstract
{Classical nova explosions occur on the surface of an accreting white dwarf in a binary
system. After ejection of a fraction of the envelope and when the expanding shell becomes
optically thin to X-rays, a bright source of supersoft X-rays arises, powered by residual
H burning on the surface of the white dwarf. While the general picture of the nova event is well established, the details and balance of accretion and ejection processes in classical novae are still full of unknowns. The long-term balance of accreted matter is of special interest for massive accreting white dwarfs, which may be promising supernova Ia progenitor candidates.
Nova~\object{V5116~Sgr}~2005b was observed as a bright and variable supersoft X-ray source by {\it XMM-Newton} in March~2007, 610~days after 
outburst. The light curve showed a periodicity consistent with the orbital period. During one third of the orbit the luminosity was a 
factor of seven brighter than during the other two thirds of the orbital period.} 
{In the present work we aim to disentangle the X-ray spectral components of V5116~Sgr and their variability.}
{ We present the high resolution spectra obtained with {\it XMM-Newton} RGS and {\it Chandra} LETGS/HRC-S in March and August 2007.}
{The grating spectrum during the periods of high-flux
shows a typical hot white dwarf atmosphere dominated by absorption lines of N~VI and N~VII. During the low-flux periods, the spectrum is 
dominated by an atmosphere with the same temperature as during the high-flux period, but with several emission features superimposed. 
Some of the emission lines are well modeled with 
an optically thin plasma in collisional equilibrium, rich in C and N, which also explains some excess in the spectra of the high-flux
period. No velocity shifts are observed in the absorption lines, 
with an upper limit set by the spectral resolution of 500~km~s$^{-1}$, consistent with the expectation of a non-expanding atmosphere 
so late in the evolution of the post-nova.}
{}

                \keywords{novae, cataclysmic variables 
                  -- X-rays: individuals (\object{V5116 Sgr})}

        \titlerunning{The supersoft X-ray source in the post-outburst nova \object{V5116 Sgr}. I.}
        \maketitle

%

\section{Introduction}

Classical novae occur in close binary systems of the cataclysmic variable type, 
when a thermonuclear runaway results in explosive 
hydrogen burning on the accreting white dwarf (WD) \citep{jh07, sta08}. It is theoretically predicted that 
novae return to hydrostatic equilibrium after the ejection of a fraction of the envelope.
Soft X-ray emission arises in some novae in outburst
as a consequence of residual hydrogen burning on the WD surface.
As the envelope mass is depleted, the photospheric radius 
decreases at constant bolometric luminosity 
(close to the Eddington value) with an increasing effective temperature. 
This leads to a shift of the spectral energy distribution from optical through 
UV, extreme UV and finally soft X-rays, with 
the nova emitting as a supersoft source (SSS, Greiner 1996) with a 
hot WD atmosphere spectrum.
The duration of this SSS phase is related to 
the nuclear burning timescale of the remaining H-rich envelope 
and depends, among other factors, on the WD mass \citep{TT98,SH05}.
The observed typical duration is a few months (Greiner et al 2004, Pietsch et al. 2005).

\object{V5116 Sgr} was discovered as Nova Sgr 2005b on 2005 July 4.049~UT, 
with magnitude $\sim$8.0, rising to mag 7.2 on July 5.085 \citep{lil05}. 
It was a fast nova, with the time required for decline of two and three magnitudes 
from maximum being $t_2=6.5\pm1.0$~days and $t_3=20.2\pm1.9$~days, respectively \citep{dob07}.
It was classified as an Fe II class in the Williams (1992) 
classification (Williams et al. 2008).
The expansion velocity derived from a sharp P~Cygni profile
detected in a spectrum taken on July~5.099 was $\sim$1300~km~s$^{-1}$.
IR spectroscopy on July 15 showed emission lines with FWHM~$\sim$2200~km~s$^{-1}$ \citep{rus05}. 
Photometric observations obtained during 13 nights in the period August-October~2006 
show the optical light curve modulated with a period of $2.9712\pm0.0024$~h \citep{dob07},
which the authors interpret as the orbital period. They propose that the  
light curve indicates a high inclination system with an irradiation effect on the secondary star. 
The estimated distance to V5116~Sgr from the optical light curve is $11\pm3$~kpc \citep{sal08}.
A first X-ray observation with Swift/XRT (0.3--10~keV) in August 2005 yielded 
a marginal detection with 1.2($\pm$1.0)$\times10^{-3}$~counts~s$^{-1}$ \citep{nes07a}. Two years later the source had evolved into a bright SSS, first detected by {\it XMM-Newton} on 2007~March~5 \citep{sal08}, and later 
on 2007~August~7, by Swift/XRT, 
with 0.56$\pm$0.1~counts~s$^{-1}$ \citep{nes07b}. In Sala et al. (2008) we reported on the EPIC spectra and light curve. The duration of the exposure, 12.7~ks, was just enough to cover one whole orbital period. The X-ray light curve showed two distinct phases, with the flux decreasing by a factor of seven for two thirds of the orbit. The EPIC spectral shape was the same both in the high and the low-flux periods, well fit with an ONe LTE atmosphere model \citep{MV91} with the same temperature for the two phases.  A 35~ks {\it Chandra} spectrum obtained 
on 2007~August~28 confirmed the periodic variability and was fit with a 
WD atmospheric model with N$_{\rm H}=4.3\times10^{21}$~cm$^{-2}$ and $T=4.65\times10^5$~K \citep{NelOri07}.

Here we present the high resolution X-ray spectra obtained with {\it XMM-Newton} RGS and {\it Chandra} HRC-S/LETGS.

\section{Observations}

\object{V5116 Sgr} was one of the targets included in our X-ray monitoring program
of post-outburst Galactic novae with {\it XMM-Newton} \citep{her10}. 
It was observed with {\it XMM-Newton} \citep{Jan01} on 2007~March~5, 610~days after outburst
(observation ID: 0405600201). 
The exposure times were 12.7~ks for the European Photon Imaging Cameras (EPIC) 
MOS1 and MOS2 \citep{Tur01}, 8.9~ks for the EPIC-pn \citep{Str01}, 
12.9~ks for the Reflection Grating Spectrometer, RGS \citep{Her01}, 
and 9.2~ks for the Optical Monitor, OM \citep{Mas01}, used with the U~filter in place
in imaging mode.

The {\it XMM-Newton} observation was affected by solar flares, which produced
moderate background in the X-ray instruments for most of the exposure time. 
Fortunately, our target was at least a factor of ten brighter than the background.
In addition, the source spectrum is very soft and little affected by solar flares.
We therefore did not exclude any time interval of our exposures, but paid 
special attention to the background subtraction of both spectra and light curves.
Data were reduced using the {\it XMM-Newton} Science Analysis System 
(SAS~13.5.0). Standard procedures described in the SAS documentation 
(de la Calle \& Loiseau 2008, Snowden et al. 2008) were followed. 

{\it Chandra} observed \object{V5116~Sgr} on 2007~August~24, 5.5~months after the first {\it XMM-Newton} observation
(Obs.~ID.~7462). A long 35~ks exposure was obtained with the HRC-S with the LETG grating in place. The data 
were reduced following standard procedures with CIAO~4.6, using CALDB version~4.6.1.1. 

\begin{figure}
  \includegraphics[width=6.5 cm, angle=270]{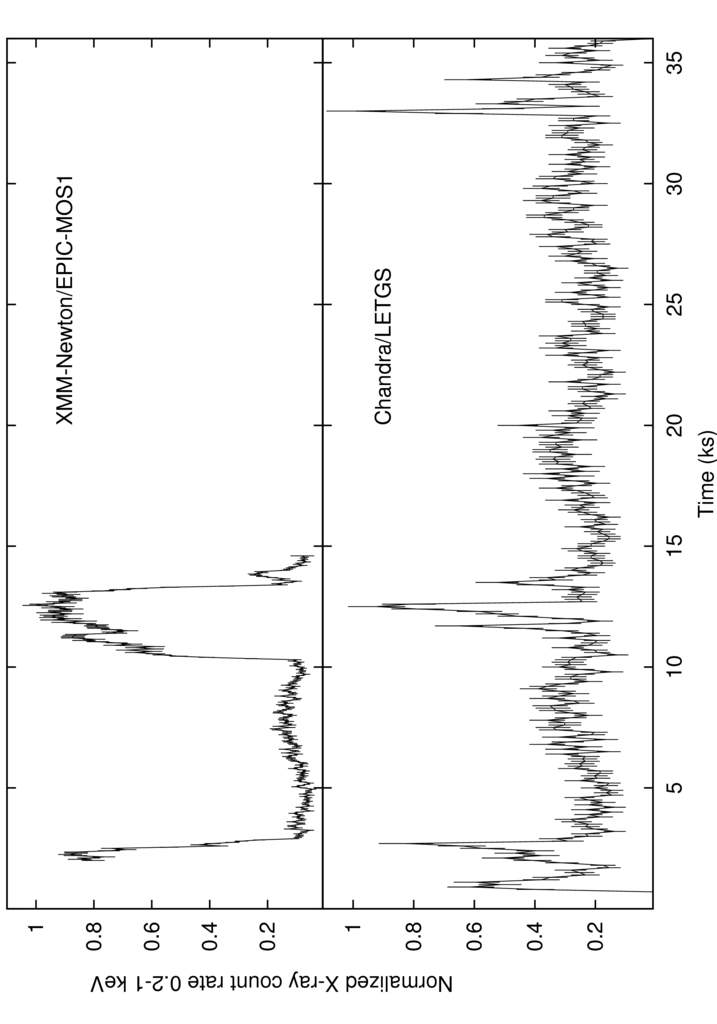}
 \caption{EPIC MOS1 (upper panel) and {\it Chandra} LETGS/HRC-S zeroth order (lower panel) normalized light curves for observations obtained in March and August 2007, respectively.}
 \label{lcs}
\end{figure}

In Sala~et~al.~(2008) we reported on the X-ray light curve and 
broadband spectra of the March~2007 {\it XMM-Newton} observation. 
The nova had evolved to a bright supersoft X-ray source.
The X-ray light curve showed abrupt decreases and increases 
of the flux (Fig.~\ref{lcs})
consistent with a periodicity of 2.97~h,
the orbital period reported by Dobrotka et al. (2008, hereafter DRL08). 
The ratio between the average count-rate in the high-flux and the low-flux is $7\pm2$, while
the ratio between the maximum and minimum count-rates is $14\pm2$. 
In the following sections we present the spectral analysis of the high-flux and the low-flux periods of the {\it XMM-Newton} observation.

The same periodicity is present in the X-ray light curve obtained more than 
five months later by {\it Chandra}. The 35~ks exposure covered four full cycles.
The {\it Chandra} X-ray light curve, first presented in Orio (2012), shows a behavior similar to the one discovered with {\it XMM-Newton}. 
Flares repeat at the orbital periodicity for three of the cycles, but periods of high-flux emission are much shorter and
the expected flare in the third observed cycle is missing (see Fig. \ref{lcs}). 
The ratio between the maximum and the minimum count rate is $4.2\pm0.5 $.
The total signal accumulated during the flares is not sufficient for a detailed spectral analysis, 
so for the present work we have limited our spectral analysis of the {\it Chandra} data to the low-flux periods (i.e., excluding the flares.)

In a second XMM-Newton observation, obtained in March 2009, V5116~Sgr was detected
as a weak X-ray source, with a 0.2--10~keV flux in the range $(5-8)\times10^{-14}\rm{erg}~\rm{s}^{-1}\rm{cm}^{-2}$. 
At 10~kpc, that is a luminosity of $(3-7)\times10^{32}\rm{erg}~\rm{s}^{-1}$. 
The source was too faint for a detailed spectral analysis, but the spectral distribution was certainly
harder than in the 2007 observation \citep{sal10}.
The phase solution for the X-ray and optical light curves obtained with {\it XMM-Newton} in August 2007, 
plus the long-term evolution of the SSS flux and the light curves of the source after turn-off of the SSS will be presented in a second accompanying paper.

\section{Revisiting the EPIC spectra with a pile-up correction}

The {\it XMM-Newton}/EPIC spectra obtained in March 2007 were presented in Sala et al. (2008). 
The spectra were soft, consistent with residual hydrogen burning in the white dwarf envelope. 
A fit with an ONe LTE white dwarf atmosphere model showed that the temperature was the same both in the low and the high 
flux periods, ruling out an intrinsic variation of the X-ray source as the origin of the flux changes.
They speculated that the X-ray light curve may result from a partial eclipse by an asymmetric structure in the  
accretion disk in a high inclination system.

The EPIC spectra were, however, severely affected by pile-up. Following Jetwa~et~al.(2015),
for our observation, with a count-rate of 10--65~cps in large window mode
(corresponding to 0.3--1.5 counts per frame), we estimate a flux loss is 5--24\%,
and a spectral distortion of 2--9\% due to pile-up.
Without a pile-up correction tool, 
Sala et al. (2008) applied the standard procedure of 
excluding the central part of the PSF to exclude the piled-up pixels during spectra extraction.   
While this could provide a good approximation for the bright, soft component of the spectrum, 
a study of any possible hard component above 0.7~keV was impossible because any faint component in that energy
range would be contaminated by the piled-up photons of the peak of the soft component (around 0.4--0.5~keV) 
wrongly detected as 0.8--1.0~keV photons.

Recently, a new option has been included in the SAS response matrix generation tool, \emph{rmfgen},
to correct for the distortion in the energy distribution caused by the pile-up of photons within a single frame. 
It does this by generating a redistribution matrix that attempts to compensate for the pile-up effects, 
which are calculated from the frequency and spectrum of the incoming photons.
We have thus applied the pile-up correction to analyze the EPIC-pn spectrum during the low-flux period (with total count-rate of about 
15~counts~s$^{-1}$).

\begin{figure}
\includegraphics[width=10cm]{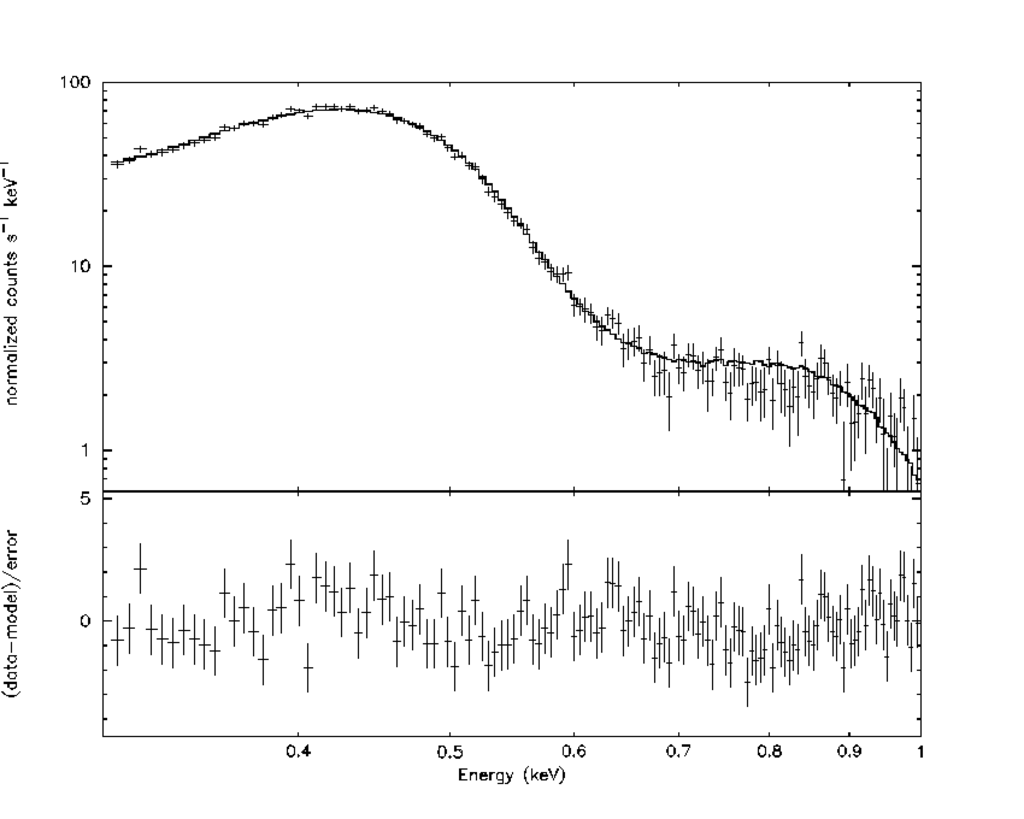}
\caption{EPIC-pn spectrum of \object{V5116~Sgr} during low-flux periods in March 2007, fit with an absorbed blackbody. 
Pile-up correction is applied to the response matrix and can be seen by good reproduction of the emission above 0.7~keV by the model. }
\label{pn_spec}
\end{figure}

Figure~\ref{pn_spec} shows the EPIC-pn spectrum during the low-flux period in March 2007. The pile-up correction is applied only to the 
response matrix, so the data themselves are not corrected for pile-up. A fit with the TBabs absorption model, with $N_{\text H}=8.7(\pm 0.1)\times 10^{21} \text{cm}^{-2}$,
and a blackbody with {\it T}$=2.1(\pm 0.2)\times 10^{5}$~K provides a good fit when folded through the new, pile-up corrected response matrix.
The ONe white dwarf LTE atmosphere model used in Sala~et~al.~(2008) also provides a good fit to the pile-up corrected EPIC-pn spectrum,
while the CO atmosphere model fails to fit the EPIC spectrum in the same way as shown in Fig. 2 of Sala~et~al.~(2008).
It is well known that fitting a blackbody to 
supersoft spectra of novae provide unphysical results (low effective temperatures with super Eddington luminosities, compensated by high absorbing hydrogen column). 
We leave the detailed spectral analysis to be presented with the RGS spectra in the following sections, but the EPIC-pn data good fit with a blackbody 
and the pile-up correction provides a test for a possible hard component, continuum or emission lines. In view of Fig.~\ref{pn_spec}, we are convinced that
the emission above 0.7~keV can fully be explained by pile-up and that no hard component is present.
It would be interesting to determine an upper limit for the flux of any possible 
component in the energy band 1--10~keV. To establish an upper limit for the flux is however dangerous because (1) it is highly model dependent and (2) any systematic 
errors in the pile-up correction would severely affect the result. The count-rate cannot be used for comparison with later observations with the same instrument, 
because the data themselves are not pile-up corrected.
Unfortunately, the correction works properly only for moderate count-rates, largely exceeded by the observed EPIC-pn count-rate 
during the high-flux periods (55~counts~s$^{-1}$) of our 2007 observation.

\section{The high resolution spectra}

\begin{figure*}
 \centering
\includegraphics[width=100mm,angle=-90]{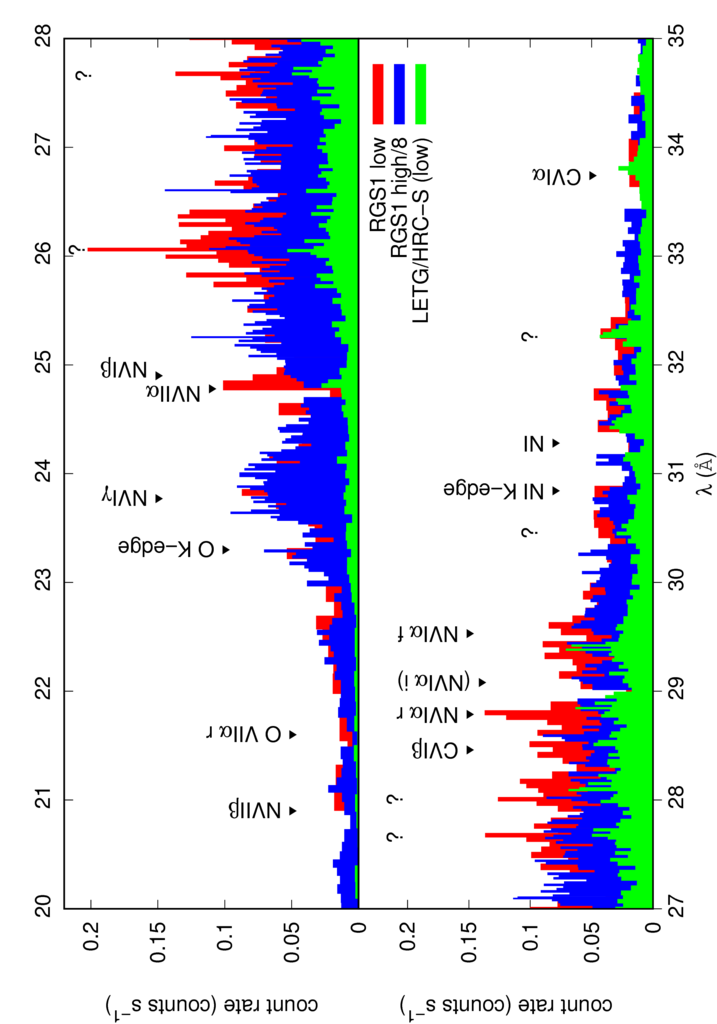}
\caption{{\it XMM-Newton}/RGS1 spectra of \object{V5116~Sgr} in high (blue) and low (red) flux periods in March 2007, and 
{\it Chandra} LETG/HRC-S excluding the flares (green), obtained in August 2007. The count-rate of the high-flux period of the RGS1 data has been downscaled by a factor of eight for ease of comparison.}
\label{spec2007}
\end{figure*}

We have extracted the high and low state RGS spectra running the SAS {\it rgsproc} 
processing chain from the same time periods as for the EPIC spectra 
given in Sala~et~al.~(2008). 
The extracted RGS spectra for the high and low states are shown in
Fig.~\ref{spec2007}. The high state spectrum is dominated by a continuum with absorption lines. 
The low state spectrum shows a number of emission lines, superimposed onto a continuum with some absorption lines. 

This pattern supports the scenario of the partial eclipse of the white dwarf: during low-flux periods, when almost 90\% of the bright X-ray source
is occulted to the observer, emission lines of the
circumstellar material become visible; during high-flux periods,
the bright SSS outshines the emission lines from the ejecta that were visible in the low-flux period, while some of the same species
located in the circumstellar material, now in the line-of-sight
to the white dwarf, appear in absorption.
Ness~et~al.~(2013) found that most super-soft high-resolution spectra of novae and conventional SS sources were either continuum dominated, with some absorption lines (SSa),
or line-emission dominated (SSe). V5116 Sgr is one of the few sources switching between the two classifications, as shown in their Figs.~5~and~6.
Sections~\ref{sect_high} and~\ref{sect_low} show the modeling of the continuum during the high-flux state plus the emission-line component during the low-flux state.
It will be shown that the high-flux state spectrum is compatible with the presence of an emission-line component as seen in the low-state spectrum, overshined
by the non-occulted bright continuum source during the high-flux state.

\subsection{Evolution of the emission features during the {\it XMM-Newton} observation}

\begin{figure*}
\centering
\includegraphics[width=145mm,angle=0]{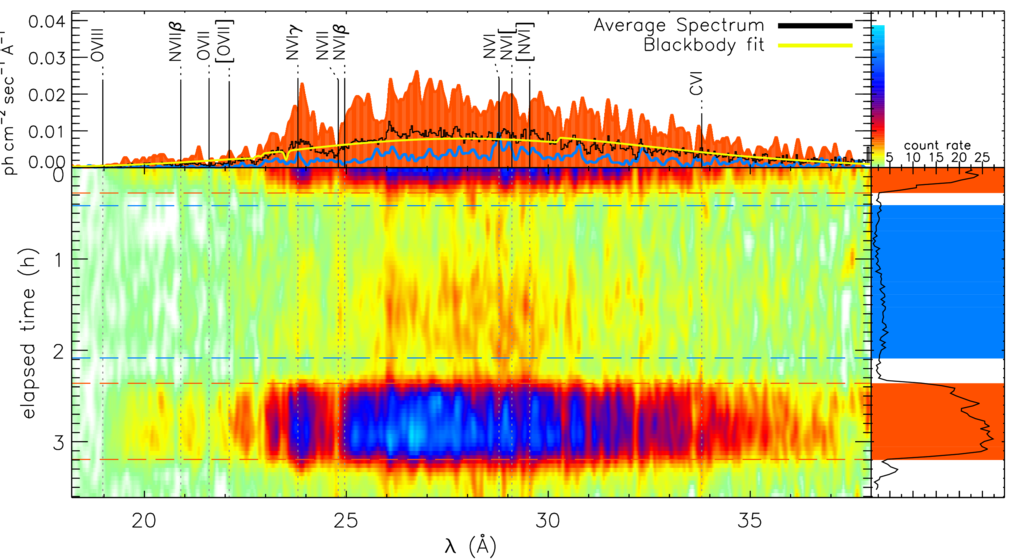}
\caption{{\it XMM-Newton} RGS time map. Evolution of the emission features during the {\it XMM-Newton} observation in March 2007.
Colors in the central panel correspond to photon fluxes, with the color key given in top right box, along the Y-axis spectrum.}
\label{timemap}
\end{figure*}

\begin{figure*}
\centering
\includegraphics[width=150mm,angle=0]{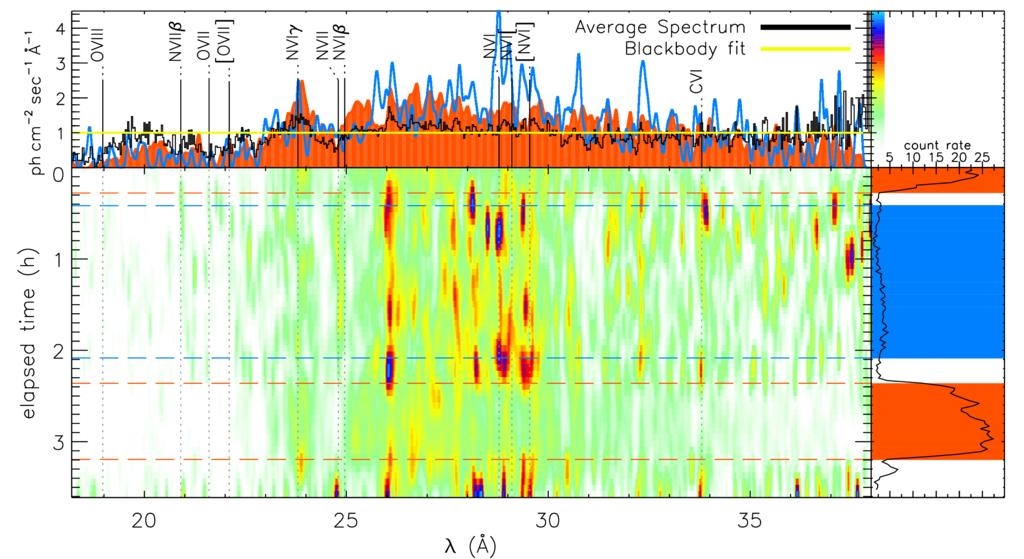}
\caption{As in Fig. \ref{timemap}, but residuals with respect to the blackbody best fit are shown.}
\label{timemap_bbnorm}
\end{figure*}

\begin{figure*}
\centering
\includegraphics[width=150mm,angle=0]{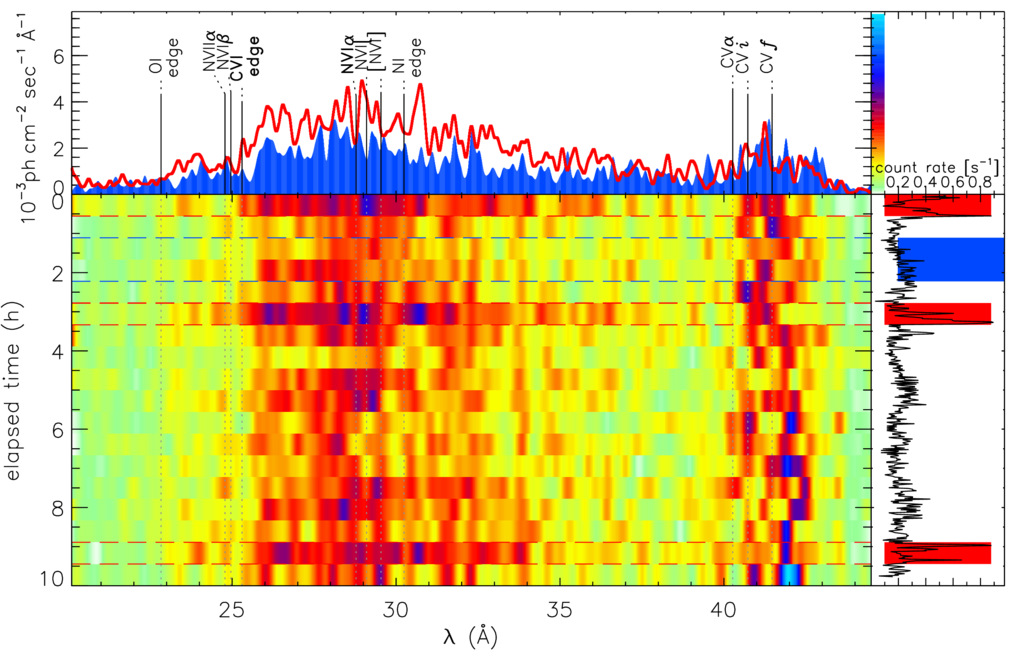}
\caption{As in Fig. \ref{timemap} for the {\it Chandra} observation in August 2007.}
\label{timemap_chandra}
\end{figure*}

We have searched for variability of the emission features detected in the low-flux spectra during the {\it XMM-Newton} observation. 
Figs.~\ref{timemap},~\ref{timemap_bbnorm}, and~\ref{timemap_chandra} show the dynamic energy spectra obtained during the {\it XMM-Newton} and the {\it Chandra} observations. 
The upper panel shows the spectrum during high- and low-flux periods. The average spectrum and the best blackbody fit are also shown in the upper panel. 
The main central panel shows the energy spectra, in 500~s bins for {\it XMM-Newton} and 2000~s bins for {\it Chandra}.
The presence and evolution of the emission features is clearer in Fig.~\ref{timemap_bbnorm}, where each 
spectrum has been divided by a best-fit blackbody fit to bring out the contributions from lines. 
It is evident that the emission lines appear mainly during the low-flux phase, and their intensity is variable in time.
In particular, some emission features are brighter at the minima of flux, just after and before the high-flux states.
During the low-flux state, the light curve shows a modulation. This modulation is also present in the low-flux states observed in August 2007 with Chandra (Fig.~\ref{lcs}).
The dynamic spectrum in Fig.~\ref{timemap} indicates that this modulation is dominated by the emission line component, rather than an intrinsic variation of the continuum source.

\subsection{The high-flux spectra: white dwarf atmosphere models}
\label{sect_high}

The spectrum during the high-flux periods, clearly dominated by a continuum with absorption lines, corresponds to the emission of the hot white dwarf. 
The LTE atmosphere models used for the EPIC spectra in Sala~et~al.~(2008) lack the details of absorption features to reproduce the line-rich high resolution RGS spectra. We fit the high-flux spectrum with the T\"ubingen NLTE atmosphere models (fluxed spectra TMAF, from the package TMAP of NLTE models, Rauch 2003, Rauch~et~al. 2010).
A grid of TMAP white dwarf atmosphere models for ten different abundances is available for use with 
XSPEC\footnote{http://astro.uni-tuebingen.de/$\sim$rauch/TMAF/ flux\_HHeCNONeMgSiS\_gen.html}.
All the TMAP models are computed for log~{\emph g}~=~9. This represents a limitation in the possible fitting results, which may lead to 
unrealistic white dwarf masses. We are cautious about this point and have checked that our results are consistent with acceptable white dwarf properties (see Sect.~\ref{discussion}).
The atmosphere model is modified with the TBabs absorption model \citep{wilms2000} to take into account the interstellar or possible intrinsic absorption.
The first fits to the RGS data showed a poor fit to the region of the oxygen K-edge.
The results improve with the use of the improved package for the TBabs model\footnote{http://pulsar.sternwarte.uni-erlangen.de/wilms/ research/tbabs/index.html}.
We use in particular the TBabs\_feo option, with free oxygen and iron abundances for the absorption, with abundances of other elements fixed to the interstellar abundances given by Wilms~et~al.~(2000).
The fit is insensitive to the iron abundance, which would contribute with its absorption edges around 17\AA, where we have no continuum.
There is however a clear oxygen overabundance in the absorbing material, as shown in the results listed in Table~\ref{tbnew_tab_rgs_mos}.

Results from simultaneous fits to RGS1 and RGS2, and EPIC-MOS1 and EPIC-MOS2, with the TBabs absorption and TMAP atmosphere models, are shown in 
Table~\ref{tbnew_tab_rgs_mos}. Simultaneous fits to EPIC-MOS data when studying the high resolution RGS spectra provide a better constraint to the absorbing hydrogen column 
affecting the soft end of the spectra. We fit our data with ten available TMAP models, each of them differing in the chemical abundances of the white dwarf atmosphere. 
For each of the ten TMAP models, we calculated the 99\% confidence range of the fitting parameters: temperature $T$ and photospheric radius $R$ for 
the white dwarf atmosphere, and oxygen abundance [O] and hydrogen column density $N_{\rm H}$ for the absorbing material (from two parameters contour plots).
All TMAP atmosphere models provide a similar fit quality, with $\chi^2_{\nu}\sim 2$, so none of them can be selected as the best-fit model for our RGS spectra. 
Taking into account the uncertainty for all the TMAP models fit to our data
we obtain the range of good fitting parameters to be $T=7.6(\pm0.4)\times10^5$~K
and $R=6(\pm3)\times 10^8$~cm from the white dwarf atmosphere model (assuming a distance to the source of 11~kpc, Sala~et~al.~2008);
and $N_{\rm{H}}=1.3(\pm0.3)\times10^{21}\rm{cm}^{-2}$ and [O]=4.4$\pm$2.2 (oxygen abundance relative to solar) for the absorbing material along the line-of-sight.
The total observed X-ray flux (0.3--1.0 keV) is $1.4(\pm0.2)\times 10^{-10}\rm{erg}~\rm{cm}^{-2}\rm{s}^{-1}$, corresponding to an unabsorbed flux of
$9.0(\pm0.5)\times 10^{-10}\rm{erg}~\rm{cm}^{-2}\rm{s}^{-1}$.
 
The observed $N_{\rm{H}}$ is consistent with the average interstellar absorption toward the source, $1.34\times 10^{21} \rm{cm}^{-2}$ (Kalberla et al. 2005).
Photometry at maximum of the nova outburst indicated $B-V=+0.48$ (Gilmore \& Kilmartin 2005), and for novae at maximum, intrinsic $B-V=0.23\pm0.06$ (van den Bergh \& Younger 1987). This implies
$A_{V}=3.1\,E_{B-V}=0.8\pm0.2$. The $N_{\rm H}$ obtained from our X-ray spectral fits indicates $A_{V}=0.7$ (using $N_{\rm H}=5.9\times10^{21}E_{B-V}\,\mbox{cm}^{-2}$, Zombeck 2007),
consistent with the value obtained from the observed colors. While the hydrogen column is perfectly compatible with the interstellar value in the line of sight, we find an enhanced oxygen abundance 
in the absorbing material between the white dwarf photosphere and the observer. This is most probably located in the circumstellar material 
around the site of the nova event.

\begin{figure*}
\centering
\sidecaption
\includegraphics[width=90mm,angle=-90]{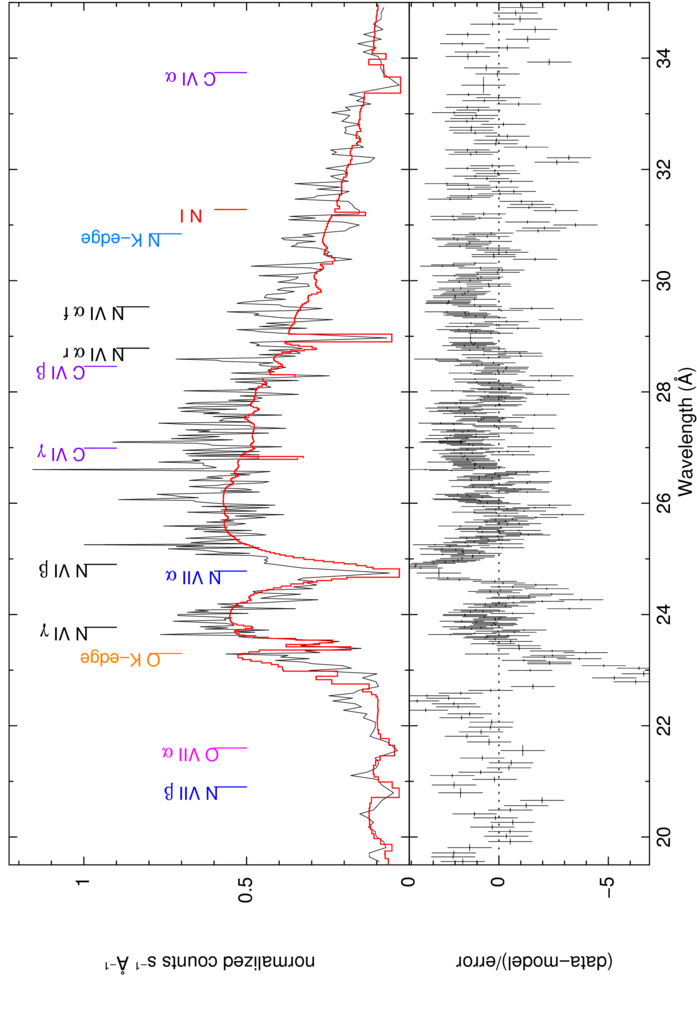}
\caption{RGS1 spectrum during the high-flux state (black), fit with the TBabs and TMAF atmosphere 003 models (red).}
\label{high_model003}
\end{figure*}

The fit to the high-flux RGS spectrum with the 003 model is shown in Fig. \ref{high_model003}. 
All TMAP atmosphere models provide poor fits around some particular absorption features.
Even though the new TBabs model improves the fit around the 
O~K-edge at 23~\AA, some details are not well reproduced by the model. This is a complex feature including O~I, O~II and O~III 
absorption lines from cool and warm interstellar and/or circumstellar material. Interstellar absorption oxygen lines have been previously observed 
in the RGS spectra of other sources, including novae \citep{nes09} and distinguished from the instrumental components around the oxygen K-edge, 
at 23.05~\AA\ and 23.35~\AA\ \citep{cor03}. O~I, O~II and O~III absorption lines have also been detected
in the high resolution {\it Chandra}/HETGS spectra of seven X-ray binaries \citep{jue04}, as part of a study 
of the structure of the oxygen K-edge originated in the interstellar and circumstellar environment of the X-ray sources.

\begin{figure*}
\centering
\includegraphics[width=90mm,angle=-90]{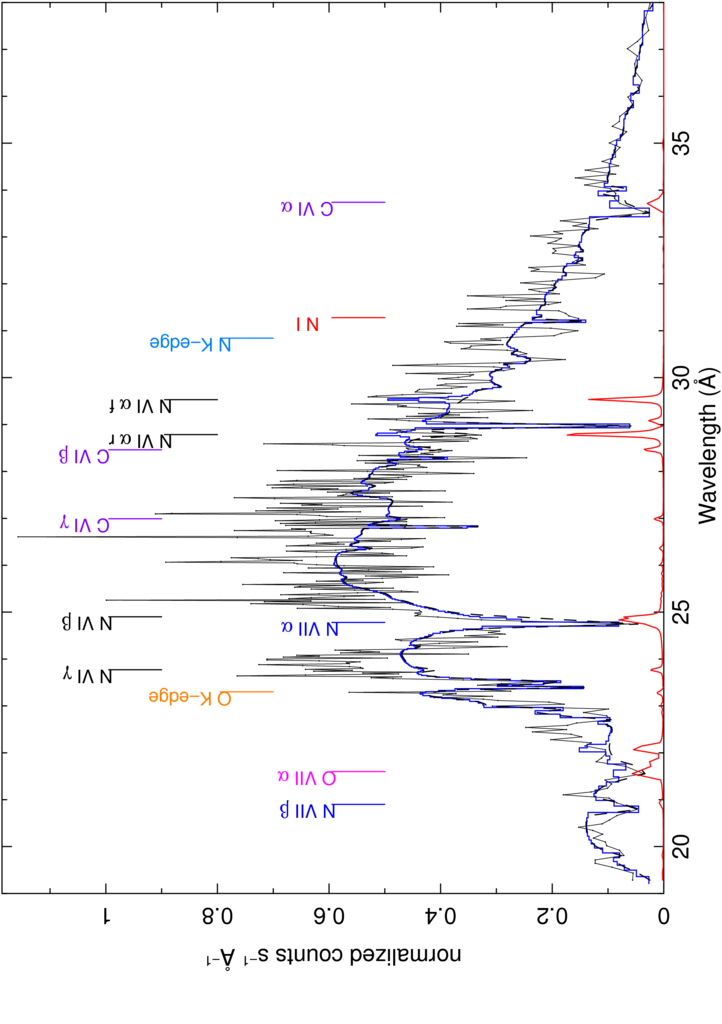}
\caption{RGS1 spectrum during the high-flux state (black), fit with the TBabs, TMAF atmosphere model 008 (green), and VAPEC plasma model (red). The total model is shown in blue. 
The VAPEC plasma model is taken with the same parameters as the best fit for the low-flux spectrum, and contributes to a better fit of the N~VI~$\beta$~line.}
\label{high_spec_vapec}
\end{figure*}

Other features poorly reproduced by the atmosphere models are the N~VII~Ly~$\alpha$~line at 24.8~\AA, the features 
around 31~\AA\ (possibly the NI~K-edge), 
and the features at 32.2~\AA\ (also present and unidentified in other novae, see Table~\ref{ulines}), and around 35~\AA.
The excess around the N~VII~Ly~$\alpha$~line at 24.8~\AA\ can be explained by the emission line component clearly detected in the low-flux emission period 
(see next section and Fig. \ref{high_spec_vapec}.)

\begin{table*}
\caption{Simultaneous fits to RGS1/2 and MOS1/2 high-state spectra in March 2007 
with Rauch (2003) atmosphere models, with the TBabs, free oxygen abundance in the absorbing material, 
and free $N_H$; radius and luminosity assuming \textit{D}= 11~kpc.
Errors quoted correspond to 99\% confidence range.}
\centering
\begin{tabular}{c c c c c c c c }
\hline\hline
Model & $T_{\text{eff}}$   &   \textit{R} & \textit{L}$_{bol}$ & [O]$^1_{TBabs}$ & $N_\text{H}$ \\ 
      &  ($10^5$K)        &   ($10^8$ cm)   &  ($10^{37}$ erg s$^{-1}$)   &               & ($10^{21}$cm$^{-2}$) \\

\hline
\hline

003     &   7.7$\pm$0.1    &    5.0$\pm$1.4  & ~6$\pm$2  & 3.8$\pm$0.8   & 1.2$\pm$0.1  \\
004     &   7.6$\pm$0.1    &    5.4$\pm$1.5  & ~7$\pm$2  & 3.5$\pm$0.8   & 1.3$\pm$0.1  \\
005     &   7.5$\pm$0.1    &    6.1$\pm$1.7  & ~8$\pm$3  & 3.0$\pm$0.8   & 1.4$\pm$0.1  \\
006     &   7.3$\pm$0.1    &    6.7$\pm$1.9  & ~9$\pm$3  & 3.2$\pm$0.7   & 1.5$\pm$0.1  \\
007     &   7.3$\pm$0.1    &    6.8$\pm$1.9  & ~9$\pm$3  & 3.3$\pm$0.7   & 1.5$\pm$0.1  \\
008     &   7.3$\pm$0.1    &    6.7$\pm$1.9  & ~9$\pm$3  & 3.6$\pm$0.6   & 1.5$\pm$0.1  \\
009     &   7.3$\pm$0.1    &    6.8$\pm$1.9  & ~9$\pm$3  & 3.8$\pm$0.6   & 1.5$\pm$0.1  \\
010     &   7.3$\pm$0.1    &    6.9$\pm$2.0  & 10$\pm$3  & 4.0$\pm$0.6   & 1.5$\pm$0.1  \\
011     &   7.3$\pm$0.1    &    7.0$\pm$2.0  & 10$\pm$3  & 4.2$\pm$0.6   & 1.5$\pm$0.1  \\
201     &   7.9$\pm$0.1    &    4.3$\pm$1.2  & ~5$\pm$2  & 5.0$\pm$1.5   & 1.1$\pm$0.1  \\

\hline
\hline
\end{tabular}
\label{tbnew_tab_rgs_mos}
\begin{list}{}{}
\item[$^{\mathrm{1}}$] Abundance relative to solar. 
\end{list}
\end{table*}

\subsection{Low-flux spectrum: atmosphere plus emission lines}
\label{sect_low}

The RGS spectrum during the low-flux periods shows a strong continuum with several emission lines 
superimposed. A plasma emission model (APEC or MEKAL in XSPEC) alone fails to fit 
the observed spectrum. A reasonably good fit is obtained with a TMAP atmosphere model, but 
several emission features clearly leave an excess of the data over the model. 

\begin{figure}
\includegraphics[width=60mm,angle=-90]{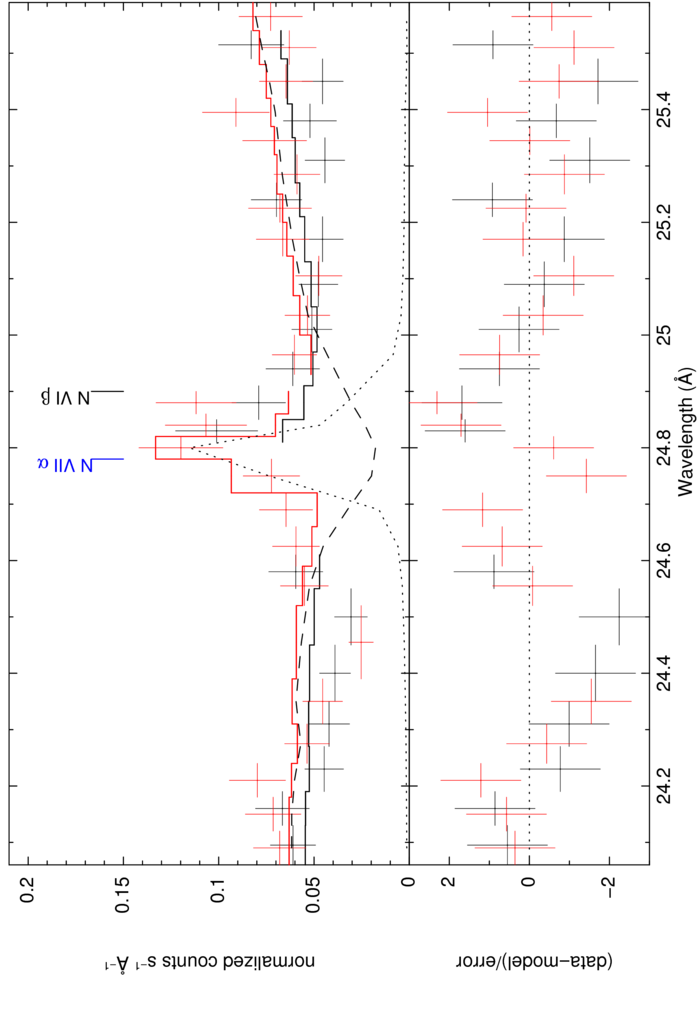}
\caption{Detail of the RGS1 (black) and RGS2 (red) spectra in low-flux state around the N~VII~$\alpha$ absorption line, modeled by the TMAP atmosphere model 008 (dashed curve). The excess in the data left by the atmosphere model could be due either to N~VI~$\beta$~line blueshifted by~$\sim$1300~km~s$^{-1}$,
or to N~VII~$\alpha$~line redshifted by around $\sim$500~km~s$^{-1}$.}
\label{detail_nvii}
\end{figure}

Table \ref{low_lines} provides the parameters of the identified emission lines above the continuum. 
Greek letters $\alpha$, $\beta$ and $\gamma$ indicate 1s-2p, 1s-3p and 1s-4p resonance transitions,
respectively, \emph{f} indicates forbidden lines in the N~VI He-like triplet (1s-2s$^3\rm{S}_1$).
We fit a Gaussian to each emission line, using only a narrow range of wavelengths around the emission line and a flat continuum.
Only for the fit of the emission line at 24.9~\AA\, which is superposed to the absorption N~VII~$\alpha$~line, the continuum is modeled by the best-fit atmosphere model instead of a flat continuum. 
Therefore we caution that the results of the fit to the emission line at 24.9~\AA\ may suffer from systematics due to the continuum model (see Fig.~\ref{detail_nvii}).
Due to the proximity to the N~VII~$\alpha$ absorption line at 24.8~\AA\, the identification of this emission line is uncertain: it could be either N~VI~$\beta$~line blueshifted by $\sim$1300~km~s$^{-1}$,
or N~VII~$\alpha$ emission line, redshifted by $\sim$500~km~s$^{-1}$.
The signal-to-noise of our data is not good enough to provide a fit without imposing some restrictions.  
We thus obtain uncertainties for the line wavelength and flux exploring the $\chi^2$ space around the best fit values. The line width is in most cases smaller than the RGS/HRC resolution.
Only for the 24.9~\AA\ emission line is the line width compatible with a value larger than the instrumental resolution, 
but no good fit can be obtained because it is superposed to the N~VII~$\alpha$ absorption line. 

\begin{figure}
\includegraphics[width=60mm,angle=-90]{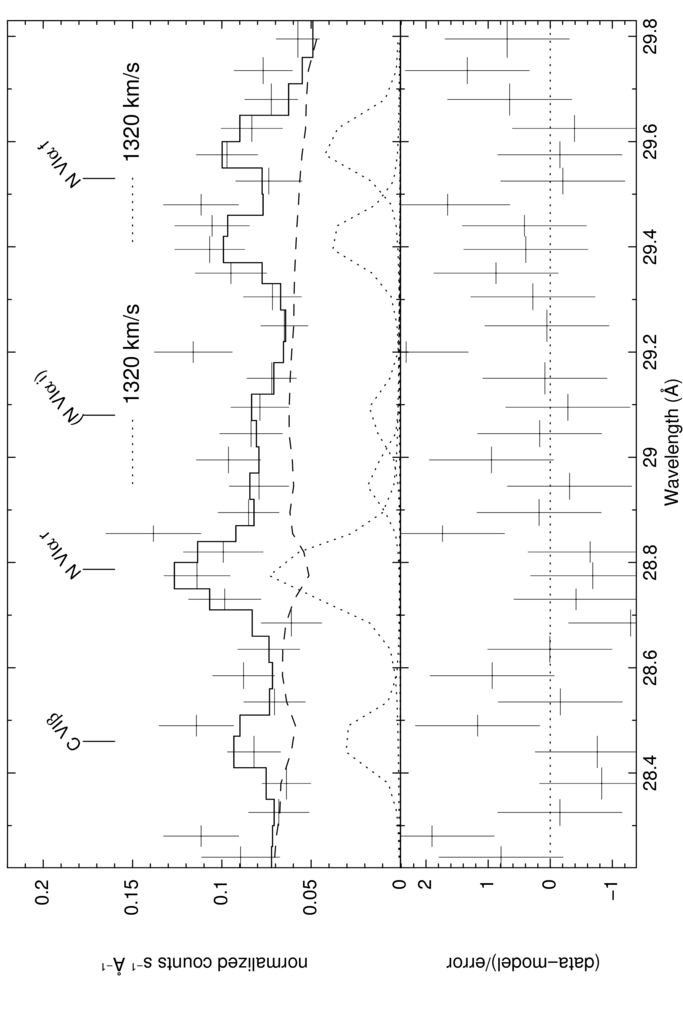}
\caption{Detail of the RGS2 spectrum in low-flux state around the N~VI triplet. Only RGS2 data are shown for clarity, but both RGS1 and RGS2 were used simultaneously for the fit.}
\label{detail_nvi_triplet}
\end{figure}

\begin{figure}
\includegraphics[width=60mm,angle=-90]{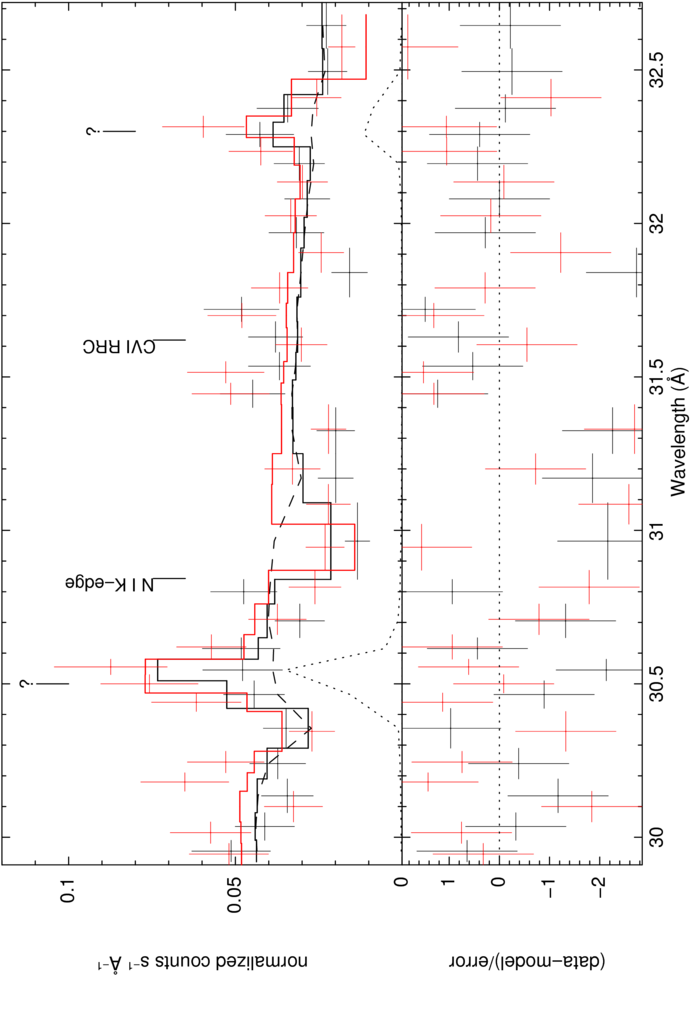}
\caption{Detail of the RGS 1 (black) and RGS 2 (red) spectra in low-flux state around 30-32\AA, with several unidentified features fit by Gaussian lines in emission (dotted curves). The NI~K absorption edge at 30.84\AA\ is not deep enough in the TBabs absorption model 
(dashed curves) and an extra absorption Gaussian is used in the total model (solid line) to fit the data in this range.}
\label{detail_unidentified}
\end{figure}

\begin{figure}
\includegraphics[width=60mm,angle=-90]{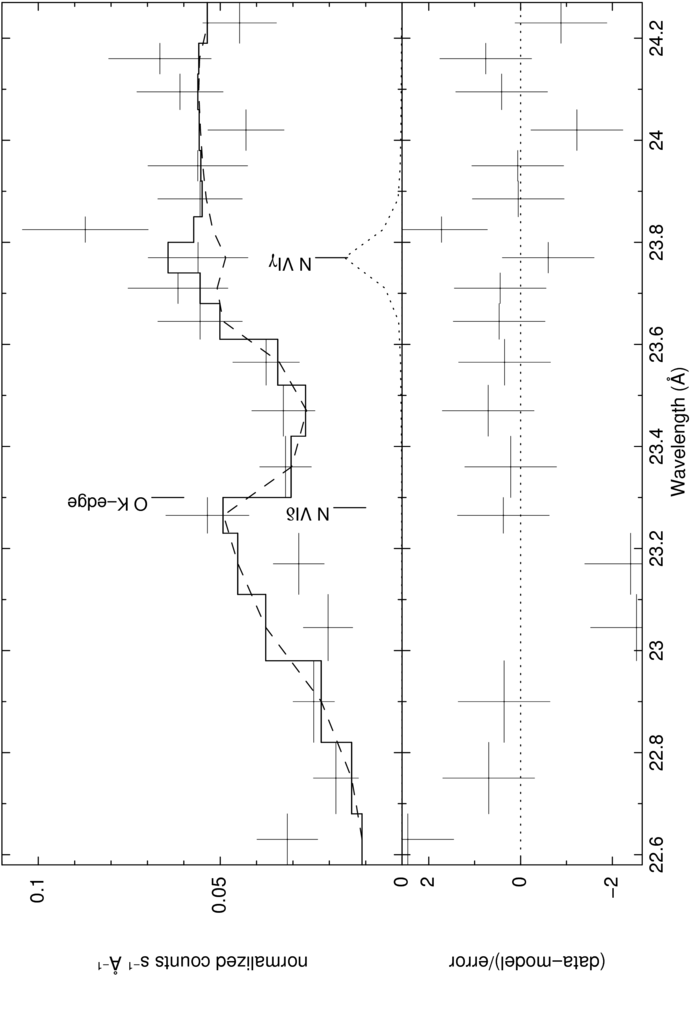}
\caption{Detail of the RGS1 spectrum in the low-flux state around the N~VI~$\gamma$ and $\delta$~lines.} 
\label{detail_nvi_gamma_delta}
\end{figure}

\begin{figure}
\includegraphics[width=60mm,angle=-90]{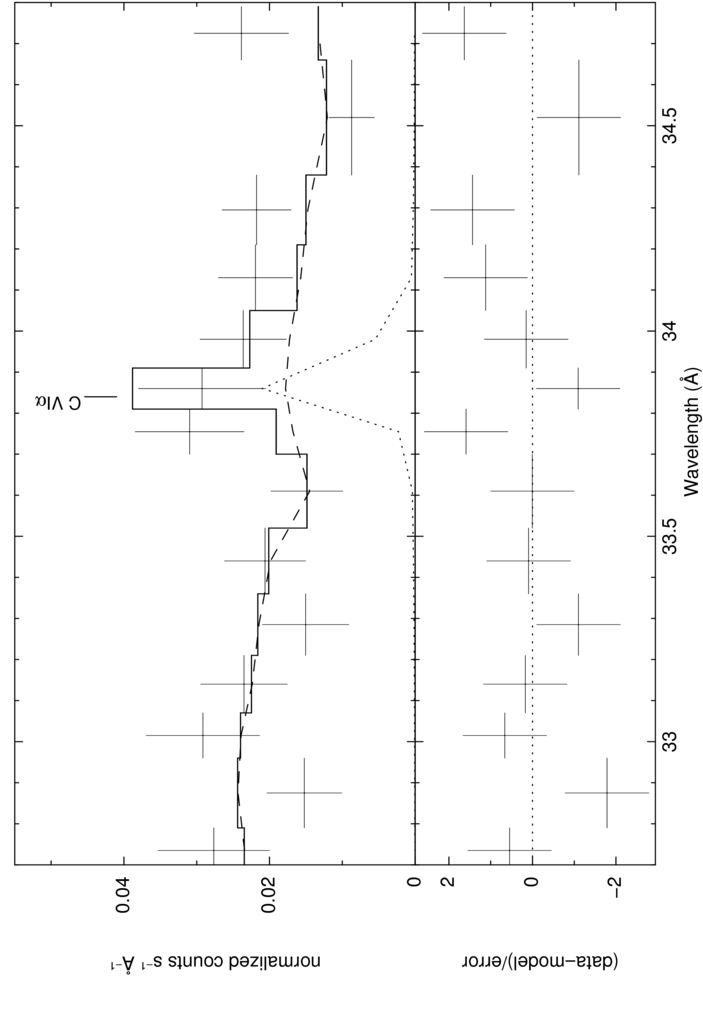}
\caption{Detail of the RGS2 spectrum in the low-flux state around the C~VI~$\alpha$~line. Only RGS1 data are shown for clarity, but both RGS1 and RGS2 were used simultaneously for the fit.}
\label{detail_cvi_alpha}
\end{figure}

Figures \ref{detail_nvii}, \ref{detail_nvi_triplet}, \ref{detail_unidentified}, \ref{detail_nvi_gamma_delta}, and \ref{detail_cvi_alpha} show details of the emission lines in the low-flux spectrum. 
Lines of C~VI, N~VI, N~VII, and O~VII are observed in the RGS spectra. In particular, the recombination \emph{r} and forbidden \emph{f} lines of the N~VI~$\alpha$ triplet are present, but not the 
inter-combination line (only upper limit found). O~VII and C~VI~lines are at their rest wavelength. 
The N~VI~$\alpha$~\emph{f}~line shows a possible second component, blueshifted by $\sim$1300~km~s$^{-1}$. This second component is not present for the \emph{r}~line, probably due to self-absorption.
The significance of the line is not high (1.7~$\sigma$ significance and F-test statistic of~3.86), but it would indicate the presence of a second, expanding gas component. 
In general, the bright atmosphere continuum with its own absorption features makes the determination of the emission line parameters uncertain. 

Strong emission features remain unidentified at 26.1~\AA, 27.7~\AA, 28.1~\AA, 30.5~\AA, and 32.2~\AA\ (Table~\ref{ulines}), which cannot be explained by collisional plasma emission. 
We note that the C~VI~$\beta$ (28.46~\AA) and C~VI~$\gamma$ (26.99~\AA)~lines, which could correspond to some of the unidentified features, cannot be as bright as suggested by the data without a 
C~VI~$\alpha$~line (33.74~\AA) being much brighter than observed. Most of these features have also been found in the high-resolution spectra of RS Oph, and in particular, the feature at
26~\AA\ seems to be common in the emission line spectra of novae in outburst, detected and unidentified in \object{V2491~Cyg}, \object{RS~Oph}~and~\object{V4743~Sgr}~(Table~\ref{ulines},
Ness~et~al.~2011).

The relation in fluxes of the three components (\emph{r, i} and \emph{f}) of He-like triplets can provide information on the conditions of the emission site. In a purely collisional plasma, 
the fluxes of \emph{i} and \emph{f} lines are similar to the flux of the recombination line (\emph{f+i$\sim$r}). The ratio G=$\frac{f+i}{r}$ is smaller than one in the presence of photoexcitation, 
while G~>~1 indicates a purely recombining plasma 
or a photoionized plasma with high column densities. Since we have only an upper limit for the flux of the intercombination line, we can only give an upper limit for G, which considering also 
the uncertainties in \emph{f} and \emph{r}, turns out to be G~<~5. In the absence of UV radiation that triggers excitations from the forbidden to the inter-combination levels, 
the three components of the He-like triplet also provide a density diagnostic, with the ratio \emph{f/i} expected to be 4.9 in 
the low density limit, log($n_{\rm{e}}$)=9.65, and \emph{f/i}~<~4.9 indicating a high density environment. In our case, the upper limit in the flux of the \emph{i} component indicates \emph{f/i}~>~1,
which indicates a density below log($n_{\rm e}$)~<~10.2.

\begin{figure*}
\centering
\includegraphics[width=100mm,angle=-90]{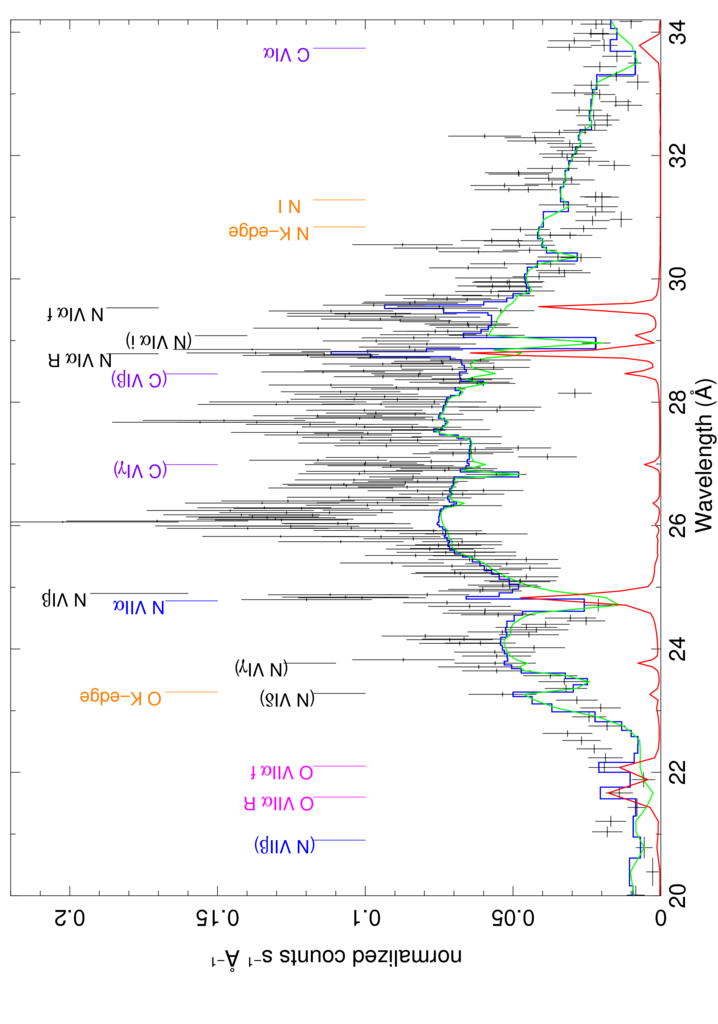}
\caption{RGS1/2 spectrum during the low-flux state (black), fit with the TBabs, TMAF atmosphere model 003 (green), and VAPEC plasma model (red) with free C, N and O abundances. 
The total model is shown in blue. Best fits are obtained with normalization of the atmosphere about a factor of eight fainter than in high-flux, an atmosphere temperature around $7\times 10^5$K, 
a plasma temperature around 0.1~keV, and poorly constrained but high overabundances of [C], [N], and [O].}
\label{low_spec_vapec}
\end{figure*}

A coherent fit to most of the observed emission lines is provided by an 
optically thin plasma in collisional ionization equilibrium (CIE), modeled using the VAPEC model in XSPEC, added to the TMAP model, with free C, N, and O abundances (see Fig.~\ref{low_spec_vapec}).  
In Table \ref{tbnew_tab_rgs_mos_low} we summarize the best-fit parameters for all available TMAP atmosphere models (different models correspond to different abundances), modified by the 
 T\"ubingen absorption model and with an additional plasma component (modeled with VAPEC in XSPEC.) 
Best fits are obtained with a normalization of the atmosphere about a factor of eight fainter than in high-flux, an atmosphere temperature around 7$\times 10^5$K, a plasma temperature around 0.1~keV, 
and high (but poorly constrained) overabundances of C, N, and O.  
As in the case of the spectrum during the high-flux period, no particular model provides a statistically  
better fit, so we consider the range of obtained parameters as the uncertainty in the parameters.
Taking into account 99\% confidence errors (from two-parameter contour plots) from fits of all atmosphere models, we obtain a temperature of the plasma component k$T_{\rm{VAPEC}}$= 0.12$\pm$0.01~keV, 
atmosphere temperature $T_{\rm{atmos}}=7.2(\pm 0.3) \times 10^5$~K, bolometric luminosity for the atmosphere $L_{\rm{atmos}}=1.2(\pm 0.4)\times 10^{37}\rm{erg}~\rm{s}^{-1}$,
oxygen abundance in the absorbing material of [O]=6$\pm$3 with respect to solar,
and a column density of $N_{\rm{H}}=1.2(\pm 0.4)\times 10^{21}\rm{cm}^{-2}$. The low sensitivity of the fit of the global spectrum to the variations in the C, N, and O
abundances prevents a good determination of the confidence ranges for their values, but typical values for the abundances required in the VAPEC component to reproduce
the emission lines observed are [C]>10, [N]>50, and [O]>10.
The total observed X-ray flux (0.3-1.0 keV) is $2.1(\pm 0.1)\times 10^{-11} \rm{erg}~\rm{cm}^{-2}\rm{s}^{-1}$.
With the measured value for the column density this corresponds to an unabsorbed total flux of $12(\pm 2)\times 10^{-11} \rm{erg}~\rm{cm}^{-2}\rm{s}^{-1}$, with only
$0.7(\pm 0.1) \times 10^{-11} \rm{erg}~\rm{cm}^{-2}\rm{s}^{-1}$ being emitted by the plasma component.

The LETG/HRC-S spectrum obtained five months later, in August 2007, with {\it Chandra} LETG/HRC-S (excluding the flares) is also dominated by a continuum with absorption 
features plus emission lines superimposed, like the low-flux states in March 2007 (see Fig. \ref{spec2007}). The global spectral distribution is softer, 
indicating a decrease of the atmosphere temperature. This is also indicated by 
the relative strength of the emission lines, with features in the range 23--28~\AA\ being fainter in the {\it Chandra} observation than in the {\it XMM-Newton} data. The details of the identified emission lines in the {\it Chandra} LETG/HRC-S spectrum are given together with 
the results for the {\it XMM-Newton} RGS in Table \ref{low_lines}. All unidentified emission lines detected in March 2007 with {\it XMM-Newton} RGS are also present in the data obtained in 
August 2007 with {\it Chandra} LETG/HRC-S (see Columns 4 and 5 in Table \ref{ulines}). 

\begin{figure*}
\centering
\includegraphics[width=100mm,angle=-90]{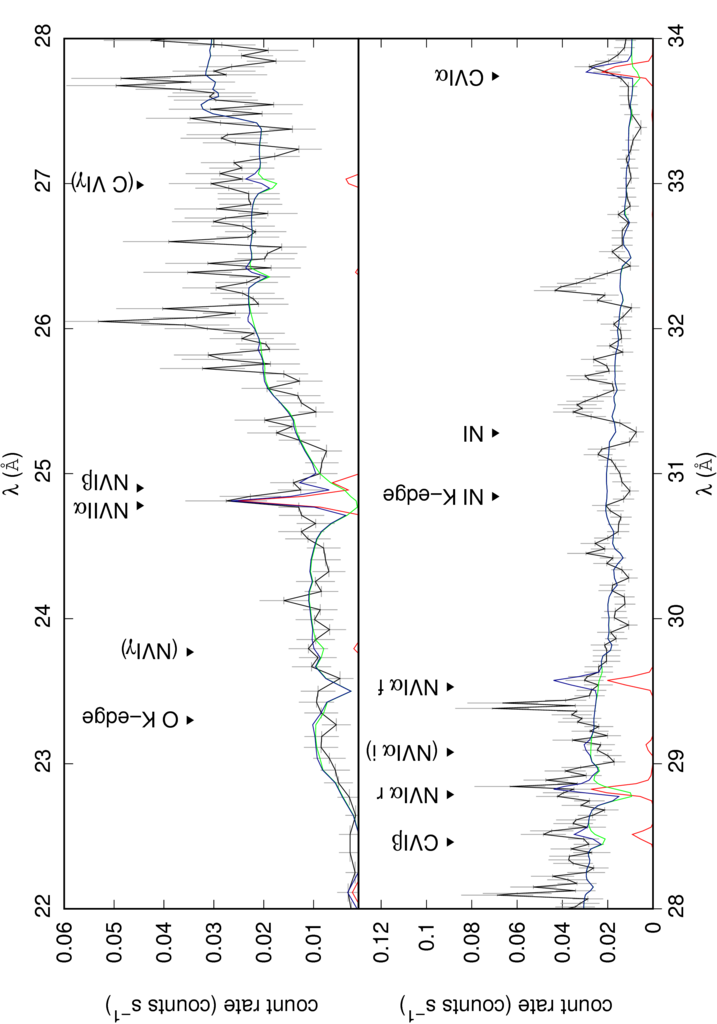}
\caption{{\it Chandra} LETG/HRC-S spectrum in August 2007, excluding the flares (black line and gray data points), fit with the  TBabs, TMAF atmosphere model 003 (green), and VAPEC plasma 
model (red) with free C, N and O abundances. The total model is shown in blue. Best fits are obtained for an atmosphere temperature around $6.8\times 10^5$K and a plasma temperature around 0.1~keV.}
\label{letg_vapec}
\end{figure*}

As for the RGS spectra, a good fit to the LETG/HRC-S spectrum is obtained with a TMAP white dwarf atmosphere model modified by the  TBabs absorption model, plus the thermal 
plasma emission modeled with VAPEC model (see~Fig.~\ref{letg_vapec}). All different TMAP atmosphere models provide similar fits, for an atmosphere 
with $T_{\rm{atmos}}=6.8(\pm 0.2) \times 10^5$~K and $L_{\rm{Bol}}=1.6(\pm 0.4) \times 10^{37}\rm{erg}~\rm{s}^{-1}$,
and a thermal plasma with k$T=0.12 \pm 0.03$~keV. Other parameters are poorly constrained in the fit.

\begin{table*}
\centering
\caption{Emission lines in RGS and LETG/HRC-S spectra during low-flux periods.}
\begin{tabular}{c c c c c c}
\hline\hline
\noalign{\smallskip}
 &    &    \multicolumn{2}{c}{RGS March 2007}  & \multicolumn{2}{c}{LETG/HRC-S August 2007}\\
ID & $\lambda_0$   & \textit{$\lambda_{obs}^1$} & \textit{F}  & {$\lambda_{obs}^1$}  & \textit{F }\\
 & (\AA)   &  (\AA) & ($10^{-13} $erg cm$^{-2} $s$^{-1}$) & (\AA) & ($10^{-13} $erg cm$^{-2} $s$^{-1}$) \\
\noalign{\smallskip}
\hline
\noalign{\smallskip}

C VI $\alpha$     &   33.74    &   33.9 $\pm 0.2$   &  9 $\pm$ 6    & 33.8 $\pm$ 0.1 & 2 $\pm$ 1 \\

\noalign{\smallskip}

C VI $\beta$      &   28.46    &   28.46 $\pm$ 0.04  &  5 $\pm$ 2  & 28.52 $\pm$ 0.04  &  5 $\pm$ 4 \\

\noalign{\smallskip}

C VI $\gamma$     &   26.99    &   -  &  < 6    &  - &  < 5.0 \\

\noalign{\smallskip}

N VI r            &    28.78   &   28.79 $\pm 0.03$ &   13 $\pm$ 5 &  28.85 $\pm$ 0.07  & 1.2 $\pm$ 0.7  \\
N VI i            &    29.08   &     -              &  < 6         &  -                 &  < 4         \\
N VI f            &    29.54   &   29.41 $\pm 0.07$ &  9 $\pm$ 5   &  29.39 $\pm$ 0.04  & 1.2 $\pm$ 0.7 \\
                  &            &   29.59 $\pm 0.07$ &  9 $\pm$ 7   &  (29.6)            &  <0.9  \\

\noalign{\smallskip}

N VI $\beta$ / N VII $\alpha$    &    24.90 / 24.78   &   24.82 $\pm$ 0.02 &  14 $\pm$3   & 24.82 $\pm$ 0.1     &  6 $^{+16}_{-3}$ \\

\noalign{\smallskip}

N VI $\gamma$     &    23.77   &     -    &  < 3  &  -    &  < 3 \\





\noalign{\smallskip}


\noalign{\smallskip}
\hline
\noalign{\smallskip}
\end{tabular}
\label{low_lines}
\begin{list}{}{}
\item[$^{\mathrm{1}}$] Line width frozen for uncertainties determination to $\sigma$=70 m\AA .
Lines fit with Gaussian over a flat continuum for a narrow wavelength range around the line of interest, except for the line at 24.9~\AA\ (see text and Fig.\ref{detail_nvii}).
\item[$^{\mathrm{2}}$] Uncertainties within 90\% confidence range. 
\end{list}
\end{table*}

\begin{table*}
\centering
\caption{Unidentified emission lines in RGS and LETG/HRC-S spectra of \object{V5116 Sgr} during low-flux periods and in other novae.}
\begin{tabular}{c c c c c c c c}
\hline\hline
\noalign{\smallskip}
& \multicolumn{2}{c}{RGS March 2007}  & \multicolumn{2}{c}{LETG/HRC-S August 2007} & \object{V2491 Cyg}$^{2}$ &  \object{V4743 Sgr}$^{2}$ &  \object{RS Oph}$^{2}$  \\
$\lambda^1$ & $\lambda_{obs}$ & \textit{F}  & $\lambda_{obs}$ & \textit{F }  & $\lambda_{obs}$ & $\lambda_{obs}$  & $\lambda_{obs}$  \\
 (\AA) & (\AA) & ($10^{-13} $erg cm$^{-2} $s$^{-1}$) & (\AA) & ($10^{-13} $erg cm$^{-2} $s$^{-1}$)  & (\AA) & (\AA) & (\AA) \\
\noalign{\smallskip}
\hline
\noalign{\smallskip}
26.06  &   26.12 $\pm$ 0.06  &    6 $\pm$ 3         &  26.05 $\pm$ 0.05  & 1.3 $\pm$ 0.6          & 25.78 & 25.85 & 25.90 \\
27.71  &     27.7 $\pm$ 0.1  &    2.5 $\pm$ 2.1      &  27.68 $\pm$ 0.06  & 1.1 $\pm$ 0.9         &  -    &   -   & 27.6  \\
28.11  &   28.1 $\pm$ 0.2    &      < 3.5            &  28.1  $\pm$ 0.1   & < 1.0                 &  -    &   -   & 28.0  \\
30.42  &  30.5 $\pm$ 0.2     &    2.1 $^{+1.2}_{-1.9}$  &  30.5  $\pm$ 0.2   & 1.3 $^{+1.3}_{-0.6}$     &  -    &   -   & 30.3  \\
  -    &  32.3 $\pm$ 0.1     &    1.8 $\pm$ 1.2      &  32.3  $\pm$ 0.5   & 1.7 $\pm$ 1.2         &  -    &   -   &   -  \\
\noalign{\smallskip}
\hline
\noalign{\smallskip}
\end{tabular}
\label{ulines}
\begin{list}{}{}
\item[$^{\mathrm{1}}$] Projected, assumed rest wavelengths of unidentified lines in Ness et al (2011), assuming blueshifts of 3300 km s$^{-1}$ for V2491 Cyg, 2400 km s$^{-1}$ for V4743 Sgr, and 1200 km s$^{-1}$ for RS Oph.
\item[$^{\mathrm{2}}$] Observed wavelength from Ness et al (2011).
\end{list}
\end{table*}

\begin{table*}
  \caption{Simultaneous fits to RGS1/2 and MOS1/2 low-state spectrum with Rauch (2003) atmosphere models plus VAPEC plasma model.
Atmosphere bolometric luminosity and plasma emission measure assume a distance of 11~kpc. Errors quoted correspond to 99\% confidence range.}
\centering
\begin{tabular}{c c c c c c c c c c}
\hline\hline

Model & $T_{\text{eff}}$   &  \textit{L}$_{atmos}$ & Plasma \textit{kT} & Plasma EM        & $N_{\text{H}}$ \\ 
      &  ($10^5$K)        &   ($10^{37}$ erg~s$^{-1}$)     &  (keV)               & ($10^{56}$cm$^{-3}$) & ($10^{21}$cm$^{-2}$)  \\

\hline\hline

003 & 7.4$\pm$0.1 & 0.9$\pm$0.3 & 0.11$\pm$0.02 & 1.4$\pm$0.6 & 1.3$\pm$0.2  \\
004 & 7.3$\pm$0.1 & 1.2$\pm$0.4 & 0.12$\pm$0.01 & 1.0$\pm$0.5 & 0.2$\pm$0.2  \\
005 & 7.2$\pm$0.2 & 1.2$\pm$0.4 & 0.12$\pm$0.02 & 1.4$\pm$0.7 & 1.3$\pm$0.2  \\
006 & 7.2$\pm$0.2 & 1.3$\pm$0.4 & 0.12$\pm$0.02 & 1.4$\pm$0.7 & 1.3$\pm$0.2  \\
007 & 7.2$\pm$0.1 & 1.3$\pm$0.4 & 0.11$\pm$0.01 & 2.2$\pm$0.7 & 1.2$\pm$0.2  \\ 
008 & 7.1$\pm$0.1 & 1.4$\pm$0.4 & 0.10$\pm$0.02 & 2.0$\pm$0.6 & 1.2$\pm$0.2  \\ 
009 & 7.2$\pm$0.2 & 1.5$\pm$0.5 & 0.12$\pm$0.02 & 2.2$\pm$0.7 & 1.0$\pm$0.2  \\ 
010 & 7.1$\pm$0.1 & 1.4$\pm$0.4 & 0.10$\pm$0.02 & 2.0$\pm$0.9 & 1.0$\pm$0.2  \\ 
011 & 7.2$\pm$0.1 & 1.2$\pm$0.4 & 0.11$\pm$0.01 & 1.4$\pm$0.7 & 1.0$\pm$0.2  \\ 
201 & 7.4$\pm$0.1 & 0.7$\pm$0.3 & 0.12$\pm$0.01 & 5.0$\pm$1.5   & 1.1$\pm$0.1  \\ 

\hline
\hline
\end{tabular}
\label{tbnew_tab_rgs_mos_low}
\end{table*}

\section{Discussion and conclusions}
\label{discussion}

The high-resolution {\it XMM-Newton} and {\it Chandra} spectra of \object{V5116 Sgr}, obtained 610 and 781 days after outburst, show two components corresponding to the white 
dwarf atmosphere plus a CIE plasma. The observation obtained by {\it XMM-Newton} on day 610 switches between a high- and a low-flux state. 
During high-flux intervals, no emission line features can be identified and the spectrum appears dominated by the hot white dwarf atmosphere emission, 
with a strong blackbody-like continuum and some absorption features. 

No velocity shifts are observed in the absorption lines, with an upper limit of $\sim$1000~km~s$^{-1}$ 
set by the spectral resolution.
This is an indication that the white dwarf atmosphere is no longer expanding.
In addition, the values for the radius determined from our fits are close to the 
expected radius for the white dwarf core,  without a significant expansion of the atmosphere. 
The observation of \object{V5116 Sgr} late in the SSS phase thus provides a good test for non-expanding white dwarf 
atmosphere models, and in this case, we find that the TMAP models provide an acceptable representation of the observed spectra. This set of white
dwarf atmosphere models suffers however from the limitation of a single \emph{log g} calculated, that is, all TMAP models available for comparison 
with the observed data are calculated for a fixed \emph{log g}=~9.
We must thus confirm that the results we obtain for the observed photospheric radius do not imply an unrealistic white dwarf mass.
For a cold white dwarf, \emph{log g}=~9 would correspond to the surface 
gravity of a $\sim$1.2~M$_{\odot}$ white dwarf with a core radius of 3.4$\times 10^8$~cm \citep{egb}.
A hot, post-outburst nova could have an expanded atmosphere and thus the photospheric radius could be larger.
In our case, as discussed in Sect. \ref{sect_high}, from the high-flux spectra, we derive a radius within the range $R=(3-9)\times 10^8$~cm,
compatible with the expected radius for a \emph{log g}= 9 white dwarf. 

The TMAP atmosphere models used to fit the spectrum during the high-flux periods are available in 10 different series corresponding to different, fixed combinations of
 abundances of H, He, C, N, O, Ne, Mg, Si, and S. However, none of the series provide a statistically significant better fit than the others. 
Despite the fact that the use of LTE models for the EPIC spectra had 
indicated better fits for ONe models than for CO in Sala et al. (2008), now in view of the RGS spectra, with the full details of the soft X-ray spectra in view, 
no conclusions can be drawn regarding the abundances of the white dwarf in \object{V5116 Sgr}. The emission line component, however, indicates high overabundances in C, N, and O
with respect to solar values.

The spectrum during the low-flux period does not show any significant change in effective temperature or hydrogen absorbing column. This points to some kind of gray obscuration of the 
central source as the origin of the flux change. While the continuum spectral distribution remains constant, just a factor of eight fainter in flux, some emission features appear during the 
low-flux period. This emission line component can be modeled by a plasma in CIE, with a temperature of 0.11--0.13~keV. Since the emission lines are not obscured during the low-flux periods, the 
emission site must be located further outside in the system than the white dwarf photosphere. 
The origin site of this component would 
probably be the circumstellar shell of material ejected during the nova event, still at high temperatures after the nova outburst. We have checked that this emission line component also
could have been present during the high-flux period, almost completely overshined by the bright white dwarf atmosphere emission. This picture is supported by the fact that an excess 
present at 24.9~\AA\ in the fit with atmosphere model of the high-flux spectra is well explained by the emission feature at this wavelength with the same line flux as found in the low-flux spectra (see Fig.~\ref{high_spec_vapec}).

The presence of the white dwarf continuum during the low-flux periods (during the occultation period), can be explained by either a partial occultation of the central white dwarf, or
by scattering of the soft X-ray photons by the surrounding material. The occultation could be produced by a warped accretion disk or, most probably, by a thick rim of the accretion disk
(Sala et al., in preparation). We note that an eclipse by the secondary would produce a short decline of the soft X-rays, and that it cannot explain an occultation of the central source
during two thirds of the orbital period.  A partial occultation of the small central white dwarf by a structure of the outer accretion disk would require a fine tuning of the system parameters to 
reduce the visible surface of the central object by a factor of seven during two thirds of the orbit. It is more likely that the occulting structure covers completely the bright central white dwarf, 
and that the continuum detected during the low-flux periods corresponds to the white dwarf emission Thomson scattered by the material surrounding the system. The scenario is similar to
the case of USco (Ness~et~al.~2012, Orio~et~al.~2013). The occulter may not be homogeneous,
hence partially transparent to X-rays, leading to a small variation in the spectral shape (see Fig.~\ref{timemap}).
We note that the high-flux states in the {\it Chandra} observation are shorter than five months before, when observed by {\it XMM-Newton} (Fig.~\ref{lcs}).
In addition, the high-flux states show a variable light curve, with different patterns in each of the orbits, and even it is completely missing for one of the
orbits. This may indicate that the absorber is evolving with time, becoming denser or covering a larger area. This could correspond to the process of reformation of the accretion
disk, as observed for USco (Ness~et~al.~2012).

The details of the line emission component suggest a complex structure in the ejected shell. While the bulk of the emission line component is well modeled by a CIE plasma with a 
unique temperature and no velocity shift, the details of the N~VI He-like triplet show a possible second component for the f line (the r line is most probably self-absorbed), 
blueshifted by $\sim$1300~km~s$^{-1}$ (Fig.~\ref{detail_nvi_triplet}). 
This would indicate the presence of at least two shells of material, one at zero velocity and the second one still expanding. 
The emission line at 24.9~\AA\ could also belong to this same component if identified as N~VI~$\beta$~line blueshifted by  $\sim$1300~km~s$^{-1}$. 
However, the larger, different line width of this line compared to the lines in the N~VI~$\alpha$ triplet argue against this identification. 
The alternative identification, a redshifted N~VII~$\alpha$ emission line, would indicate the presence of a component receding at 500~km~s$^{-1}$.

The details of the NVI He-like triplet in the LETGS data (Fig.~\ref{letg_vapec}) indicate an anomalous f/r ratio, smaller than predicted by the collisional ionization model. This may indicate that the 
assumptions for the VAPEC model fail, either because the density is higher than assumed for an optically thin CIE plasma, or that there is some photoionization process taking place.

Also the presence of several unidentified features (Table~\ref{ulines}) indicates a possible extra component or process, in addition to the white dwarf 
atmosphere and the optically thin CIE plasma. The time maps (Figs.~\ref{timemap} and \ref{timemap_bbnorm}) 
show that the unidentified lines at 26~\AA\ and 28~\AA\ become stronger at different times than the N~VI triplet,
supporting the idea that the emission site of the unidentified lines is different than the thin plasma component.
The fact that most of these features are also present in the high-resolution spectra of other novae suggests that
this extra component or process is not unique to \object{V5116 Sgr}, but rather common in post-outburst super soft X-ray emitting novae.  

With its switch between high and low-flux, 
probably due to some occultation effect by an asymmetric disk in a high inclination system, the post-outburst nova V5116~Sgr offers the rare opportunity 
to observe the central white dwarf and the plasma of the ejected shell in the same source with alternate intensity. 
As found in other post-outburst novae, the emission line component shows more than one velocity shell and also several 
unidentified lines. All this demonstrates the complexity of the ejection processes in play in the nova outburst.

\begin{acknowledgements}
We thank the collaboration of Richard Saxton to the correction of the pile-up in the EPIC-pn camera,
and fruitful conversations with Jordi Jos\'e. 
This work is based on observations obtained with {\it XMM-Newton}, an ESA science mission with instruments and contributions directly funded by ESA Member States and NASA. 
G.S. acknowledges support from the Faculty of the European Space Astronomy Centre (ESAC) and the Max-Planck-Institut f\"ur extraterrestrische Physik (MPE). 
We acknowledge the support of the Spanish MINECO grants AYA2014-59084-P (G.S.) and ESP2015-66134-R (M.H.), 
and the Generalitat de Catalunya grants SGR0038/2014 (G.S.) and SGR 1458/2014 (M.H.).
\end{acknowledgements}

\end{document}